\documentclass[12pt,preprint]{aastex}

\shorttitle{Pulsation of EE~Cam}
\shortauthors{Breger, Rucinski, \& Reegen}

\begin{document}

\title{Pulsation of EE~Cam}

\author{M. Breger\affil}
\affil{Institut f\"ur Astronomie der Universit\"at Wien, 
T\"urkenschanzstr. 17, A--1180 Wien, Austria}
\email{michel.breger@univie.ac.at}

\author{S.~M.~Rucinski\affil}
\affil{Department of Astronomy and Astrophysics, 
University of Toronto David Dunlap Observatory, 
P.O.\ Box 360, Richmond Hill, Ontario, Canada}
\email{rucinski@astro.utoronto.ca}

\and

\author{P. Reegen\affil}
\affil{Institut f\"ur Astronomie der Universit\"at Wien, 
T\"urkenschanzstr. 17, A--1180 Wien, Austria}
\email{reegen@astro.univie.ac.at}

\begin{abstract}
EE~Cam is a previously little studied Delta Scuti pulsator 
with amplitudes between those of the HADS (High-Amplitude 
Delta Scuti stars) group and the average low-amplitude pulsators. 
Since the size of stellar rotation determines both which 
pulsation modes are selected by the star as well as their amplitudes, 
the star offers a great opportunity to examine the 
astrophysical connections. Extensive photometric measurements 
covering several months were carried out. 15 significant pulsation 
frequencies were extracted. The dominant mode at 4.934~cd$^{-1}$ 
was identified as a radial mode by examining the phase shifts at different 
wavelengths. Medium-dispersion spectra yielded a $v\sin i$ value of 
$40 \pm 3$ km s$^{-1}$. This shows that EE~Cam belongs to the important 
transition region between the HADS and normal Delta Scuti stars.
\end{abstract}


\section{Introduction}

The Delta~Scuti stars are common pulsators on and near the main-sequence 
situated inside the classical instability strip. They generally pulsate
with nonradial modes and small photometric 
amplitudes in the millimag range. These stars also essentially share the high 
rotational velocity of the average star of spectral type A.
However, a number of Population I stars shows dominant radial 
modes, amplitudes in excess of 0.3 mag and rotates with a projected 
rotational velocity less than 30 km s$^{-1}$.
The members of this subgroup are called HADS 
(High-Amplitude Delta Scuti stars).

For the Delta Scuti stars, there exists a strong connection between 
rotation and pulsation properties, such as amplitude. We are presently 
engaged in a program to examine the stars between the two extremes of 
amplitude (e.g., for 44~Tau, see \citealt{Antoci et al 2007, Zima et al 2007}). 
We note that with extensive data and
nonradial mode identifications the aspect angle can be determined 
accurately, so that the true rotational velocity can be obtained 
from the spectroscopic $v \sin i$ measurements.

The list of pulsators in the HIPPARCOS catalogue \citep{ESA 1997}, as 
reanalyzed for multiperiodicity by \citet{Koen 2001}, contains a number of
stars belonging to this 'intermediate' group. EE~Cam 
(HIP~27199, HD~37857) is a promising target with reported 
peak-to-peak amplitudes near 80 millimag. Its rotational velocity is 
presently unknown, but was suspected to be low because of the 
relatively large pulsational amplitude. In particular, we are 
interested in detecting another variable such as 44~Tau 
\citep{Antoci et al 2007} which has a $v\sin i$ value of only 
2 km s$^{-1}$, in order to separate the effects of true low rotational 
velocity and geometric aspect. \citet{Olson 1980} classified EE Cam as 
an F3 giant (gF3). The catalog by \citet{Nordstrom et al 2004} gives a 
temperature of 6530K and a [Fe/H] abundance of 0.06 relative to the Sun.
\citet{Koen 2001} analyzed the Hipparcos data and suggested two frequencies 
of 4.93 and 5.21 c/d, respectively.

\section{New photometry}

In 2006 and 2007, photometric observations were performed using the 
Vienna University Automatic Photoelectric Telescope 
(APT; \citealt{Strassmeier et al 1997, Granzer Reegen Strassmeier 2001}) 
located at Washington Camp, Arizona, USA. The ``Wolfgang'' telecope 
acquired altogether 304 hours of Str\"omgren $vy$ data from 
2006 February 14 to 2006 April 2 and from 2006 September 19 to 
2007 April 5. The three-star technique was employed with 
HD~35606 (C1, $V = 8\fm15$, $B-V = 0.48$, F8) and HD~32745 
(C2, $V = 8\fm21$, $B-V = 0.96$, G0) as comparison stars. 
Since for C2, long-term variability at time scales of days could 
not be excluded, the data reduction was applied solely 
relying on C1. While the three-star technique was not applied in the
final reductions, the importance of the technique in checking for possible small-amplitude
variability in the chosen comparison stars was again demonstrated.

\begin{figure*}
\begin{center}
\includegraphics*[width=150mm]{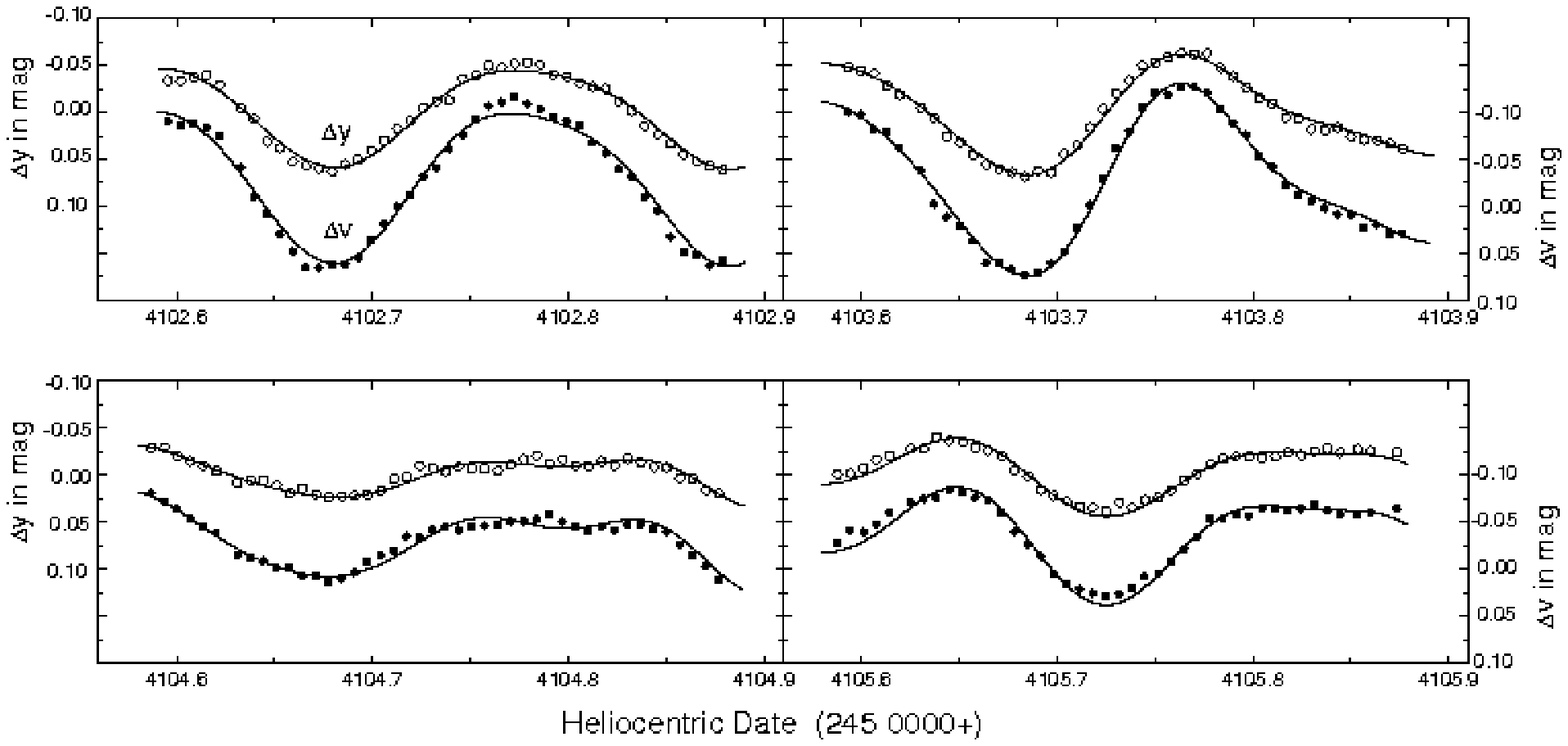}
\caption{Sample light curves of EE~Cam. The four successive nights (out
of a total of 87 nights) have typical residuals between 
the observations and the 15-frequency fits.
Both passbands $y$ and $v$ are shown.
\label{fig1}}
\end{center}
\end{figure*}

A complete list of the measurements used to 
extract the light curves is provided in Table~\ref{TABjournal}.
Typical examples of light curves of EE~Cam are shown in Fig.~\ref{fig1}
together with the multifrequency solution derived in the next section.

\section{Multiple frequency analysis}

The pulsation frequency analyses were performed with a package of computer
programs with single-frequency and multiple-frequency techniques 
(PERIOD04\footnote{The computer package to determine periodicities 
can be obtained from http://www.univie.ac.at/tops/period04.}, 
\citealt{Lenz Breger 2005}), 
which utilize Fourier as well as multiple-least-squares algorithms. 
The latter technique fits up to several hundred simultaneous sinusoidal 
variations in the magnitude domain and does not rely on sequential 
prewhitening. The amplitudes and phases of all modes/frequencies 
are determined by minimizing the residuals between the measurements 
and the fit. The frequencies can also be improved at the same time.

To decrease the noise in the power spectra, we have combined the 
measurements obtained in the $y$ and $v$ passbands. The dependence 
of the pulsation amplitude on wavelength was compensated by 
multiplying the $v$ data set by an experimentally determined 
factor of 0.64 and increasing the weight of these data points 
correspondingly. This scaling creates similar amplitudes in both passbands but 
does not falsify the power spectra. Note that different colors and data 
sets were only combined to detect new frequency peaks in the 
Fourier power spectrum and to determine the significance of 
the detection. The effects of imperfect amplitude scaling 
and small phase shifts between colors can be shown to be 
negligible for period finding. For prewhitening, separate 
solutions were obtained for each color by multiple least-square fits.
In the analysis of the Delta Scuti Network campaign data, 
we usually apply a specific statistical criterion for judging the 
reality of a newly discovered peak in the Fourier spectra, viz., 
a ratio of amplitude signal/noise = 4.0 (see \citealt{Breger et al 1993}).

Our analysis consists of a number of different steps to be repeated. 
Each step involves the computation of a Fourier analysis 
(power spectrum) from the original data or a previously 
prewhitened fit. The dominant peaks in the power spectrum were 
then examined for statistical significance and possible
effects of daily and annual aliasing. For computing new 
multifrequency solutions, the amplitudes and phases 
were computed separately for each color, so that even 
these small errors associated with combining
different colors were avoided. Note that the new 
multifrequency solutions were always computed from the 
observed (not the prewhitened) data. Because of the day-time 
and observing-season (annual) gaps, different alias 
possibilities were tried out and the fit with the lowest 
residuals selected. The resulting optimum multifrequency 
solutions were then prewhitened and the analysis repeated
while adding more and more frequencies, until the new 
peaks were no longer statistically significant.

\begin{figure}
\begin{center}
\includegraphics*[bb=51 28 505 761,width=75mm,clip]{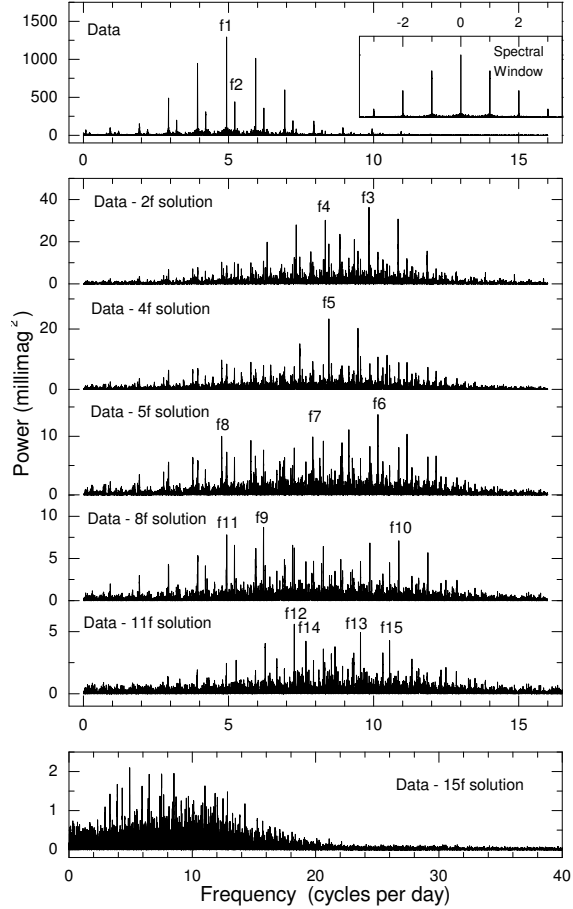}
\caption{Power spectra of EE~Cam for the 2005 and 2006 photometry. 
The panels show the power spectra after the inclusion of 
additional frequencies into the multiple-frequency solution. 
Note the 1 cd$^{-1}$ alias patterns (spectral-window 
insert of the top panel). The lowest panel clearly shows 
the power spectrum of the residuals from the 15-frequency solution:
the excess power in the 5 to 15 cd$^{-1}$ range 
shows additional pulsation frequencies and
demonstrates the rich pulsation spectrum of EE~Cam.
\label{fig2}}
\end{center}
\end{figure}

An independently performed {\sc SigSpec} analysis 
\citep{Reegen 2007}, which employs a statistical treatment 
of Discrete Fourier Transform (DFT) amplitudes that would be produced by white noise, 
provided exact consistency with the results obtained 
from the above procedure.
Fig.~\ref{fig2} shows the details of the search for the multiple frequencies, 
which are listed in Table~\ref{TABfreqs}.

The average residuals of the 15-frequency fit in the $y$ 
passband were $\pm$ 0.005 mag per single measurement, which 
is higher than the 0.003 mag expected from typical APT campaigns. 
This is  caused by the large number of presently
unresolved additional pulsation frequencies with small amplitudes, 
as revealed by the power spectrum of the residuals.

We can now compare the new multifrequency results with a 
previous result based on much fewer data: the two
frequencies of 4.93 and 5.21 cd$^{-1}$ found by \citet{Koen  
2001} are in exact agreement with our two frequencies 
showing the highest amplitudes.
 
It is possible to estimate the nonradial degree, $\ell$, 
of the dominant pulsation mode from the available photometry 
(for a recent application see \citealt{Lenz et al 2007}). 
From the $v$ and $y$ passbands we derive a value
of  $\phi(v)-\phi(y)$ = + 3.3 $\pm$ 0.4$\degr$. In this
temperature domain, such positive values generally indicate radial
pulsation. Indeed, preliminary pulsation model calculations for EE Cam
identify a unique value of $\ell$ = 0, i.e., radial pulsation.

The frequency ratio $f_1/f_2$ = 0.946. Such a frequency ratio cannot
be identified with radial modes (e.g., 
see \citealt{Suarez Garrido Goupil 2006}), so that $f_2$
has to be nonradial. We also note that $f_1$ and $f_6$ form a close
frequency pair. Such close frequencies are not unusual in Delta Scuti stars
(e.g., see \citealt{Breger Pamyatnykh 2006}). In the case of EE Cam, however,
we cannot yet exclude the possibility that the close-frequency pair
could be an artifact of strong amplitude variability of the dominant mode.

\section{Spectroscopy}

EE~Cam certainly deserves a thorough spectroscopic study.
Here we present new results which shed light on
the subject of the rotation of the star. Two spectra of EE~Cam
and one spectrum of a standard star, HD~89449 (40~Leo, F6IV),
were obtained with the David Dunlop Observatory 1.88m telescope on Feb. 14, 2006. 
For EE Cam, the UT start times were 05:15:19 (1220 s
exposure time) and 06:14:02 (1803 s exposure time), while
for HD 89449 the values were	06:52:16 and 200s.
 
\begin{figure}
\plotone{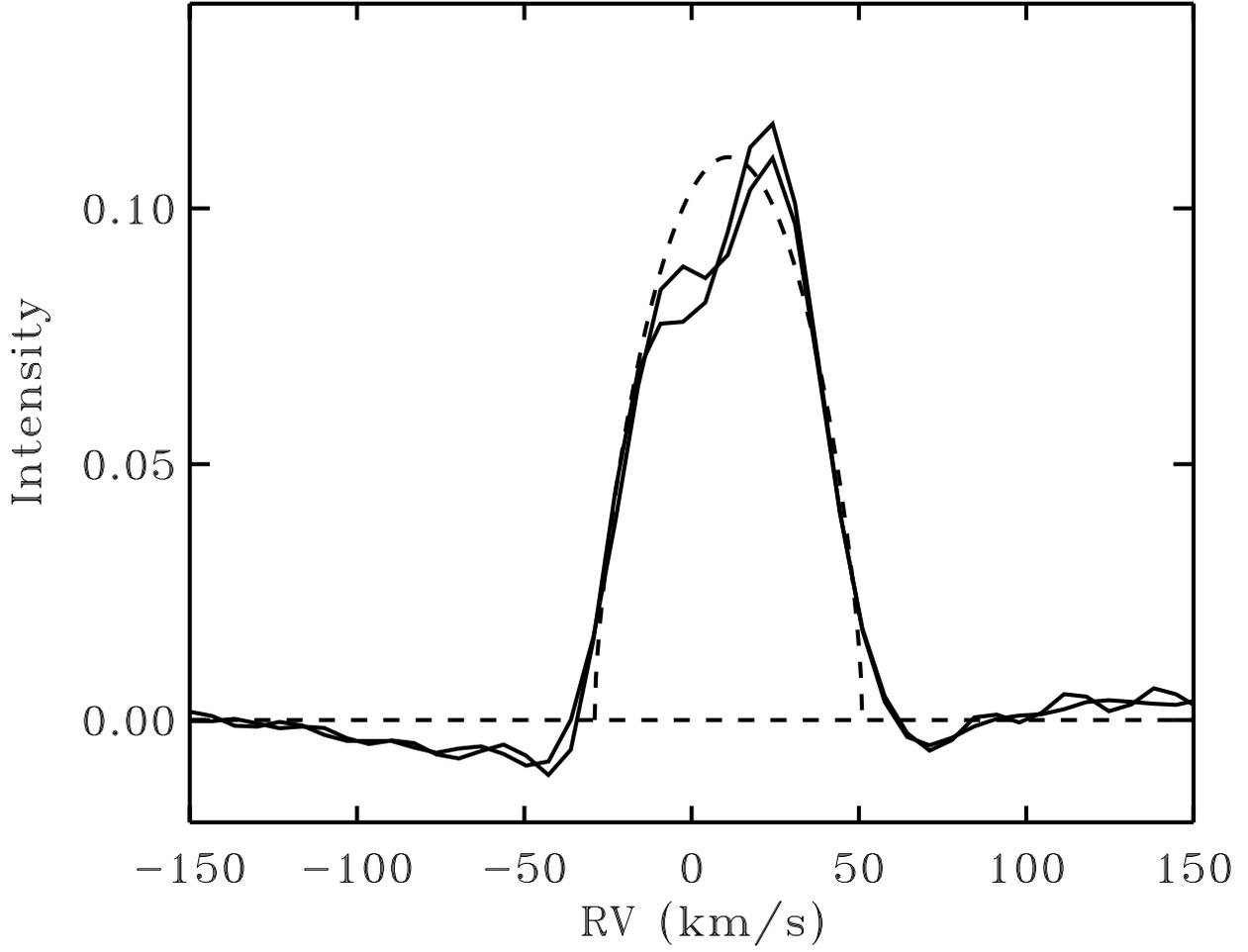}
\caption{The Broadening Functions for two spectra of EE~Cam, as
explained in the text. The rotational profile with $v \sin i = 40$
km~s$^{-1}$, centered at +11 km~s$^{-1}$, is shown by a broken line.
\label{fig3}}
\end{figure}

The spectra were centred at the Mg~I 5184 \AA\ triplet 
and covered the range 5070 to 5306 \AA\ with an effective 
spectral resolution of about 0.35 \AA. 
A full technical description is given in \citet{Rucinski 2002}.
The spectra were rectified and then processed using the broadening 
function (BF) formalism (as described in the same paper)
and subsequently improved during the 
DDO binary-star program (for the last paper of the series, 
see \citet{Pribulla et al 2007}).
The BF's were determined over the span of 61 points at 
the 6.7 km~s$^{-1}$ spacing thus covering $\pm 150$ km~s$^{-1}$. 
With such processing, a BF of a very sharp-line spectrum
has a half-width at the base of 19.5 km~s$^{-1}$.
The broken line in the figure (Figure~\ref{fig3})
shows the rotational profile calculated for the limb darkening
of $u = 0.7$ (this choice is not critical) in comparison
with the BF for EE~Cam. The BF differs from the simple
rotational profile in having a well-defined, but
unexplainable structure. The complex shape of the
BF for EE~Cam can be
interpreted as: (1) a superposition of two sharp peaks, 
possibly from two different, unresolved stars with
slightly different velocities or (2) 
one broad peak with some self-absorption, or (3) 
signatures of two shocks propagating through the atmosphere.
The most likely is the last interpretation as the
first two have no real support in what we know about the
star. If the broadening of the BF 
at the base is interpreted by rotation
of EE~Cam, then the observed half-width at the
base (45.4 and 44.8 km~s$^{-1}$ respectively) can be
used to estimate $v \sin i$. The two BF's, when corrected for
the intrinsic broadening introduced by the formalism 
and for $v \sin i \simeq 17$ of the template itself, give
$v\sin i = 40$  km~s$^{-1}$. The real uncertainty is larger
than one determined from the difference of the two spectra
and is about 3 km~s$^{-1}$
as based on results for similar stars observed in
the DDO programme.
Note that the depression of the
baseline around the BF peak is of no importance;
it is a characteristic feature for BF's of 
stars with rotationally broadenened lines as this
reflects the uncertainty of the pseudo-continuum
placement in the spectrum rectification step. 
At the time of observations, the star appeared
to have the mean heliocentric velocity of $+11 \pm 1$ 
 km~s$^{-1}$, on the assumption that the radial
velocity of the template star  
HD~89449 is $+6$ km~s$^{-1}$. 
This assumed the fit by the rotationally broadened profile,
as shown in Figure~\ref{fig3}; 
note that the standard/template star in the BF technique
is used to determine the broadening profile for the
program star as shown in this figure (and used
for the $v\sin i$ determination) as
well as the relative radial velocities.
The mean radial velocity of
EE~Cam was estimated at $+14.9$  km~s$^{-1}$ in
\citet{Nordstrom et al 2004}. 

The intensity of the BF can be used as an
indication of the spectral match of the template;
an integral of unity indicates a perfect match
and an identical spectral type (or more exactly,
an identically strong Mg~I 5184 \AA\ triplet). The two
spectra gave the BF integrals of 0.91 and 0.90 which
means that the lines are weaker than in the
F6IV standard HD~89449 indicating a spectral type
close to F5. This is confirmed by the data in the
Tycho-2 Catalog \citep{Hog et al 2000} (star number GSC~4098-123)
where the mean magnitudes give a well defined color index 
$B-V=0.427$ which correponds to the spectral type F5
on the Main Sequence.

\section{Conclusion}

Extensive photometric measurements at the millimag level 
covering several months were carried out. 
The frequency analysis has revealed 15 significant
 pulsation frequencies. The residuals show
that many additional modes in the 0 to 15 cd$^{-1}$ 
are present with small amplitudes. The
dominant pulsation at  4.934 cd$^{-1}$ was 
identified as a radial mode by examining the phase shifts
of the light curves at different wavelengths. 
The second most dominant mode at 5.214 cd$^{-1}$ was
found to be nonradial. This star, therefore, is an excellent 
example of a star showing both the properties of the HADS 
and the common small-amplitude pulsators,
in which the radial modes are either absent or very weak.

This picture of a star in the astrophysical transition region 
is supported by new measurements of the
projected rotational velocity: two medium-dispersion spectra 
yielded a $v\sin i$ value of  $40 \pm 3$ km s$^{-1}$.
This value is higher than the upper limit of 30 km s$^{-1}$ for the HADS,
but lower than the rotational velocity of the typical low-amplitude
Delta Scuti star.

Due to the relatively high amplitudes and the rich pulsation 
spectrum, EE Cam is ideal for further detailed studies of mode identification
in order to compare to asteroseismic models.

\acknowledgments

It is a pleasure to thank Heide DeBond and Jim Thomson for 
assistance with the spectroscopic observations. It is a 
pleasure to thank Patrick Lenz for computing a preliminary 
pulsation model. This investigation has been supported by the
Austrian Fonds zur F\"{o}rderung der wissenschaftlichen Forschung and 
the Natural Sciences and Engineering Council (NSERC) of Canada.

\clearpage

\clearpage

\clearpage

\clearpage
\begin{deluxetable}{rrrrr}
\tablecolumns{5}
\tablewidth{0pt}
\tablenum{1}
\tablecaption{Journal of the photometric observations of EE~Cam\label{TABjournal}}
\tablehead{
 \colhead{Start}	&
 \colhead{Length}	&
 \colhead{}		&
 \colhead{Start}	&
 \colhead{Length}	\\
 \colhead{HJD}		&
 \colhead{hours}	&
 \colhead{}		&
 \colhead{HJD}		&
 \colhead{hours}	\\
 \colhead{245 0000+}	&
 \colhead{}		&
 \colhead{}		&
 \colhead{245 0000+}	&
 \colhead{}
}
\startdata
3791.5925	&	3.30	& \hspace{10mm} &	4079.6626	&	1.78	\\
3796.6821	&	0.80	& \hspace{10mm} &	4080.6584	&	2.28	\\
3797.6170	&	1.18	& \hspace{10mm} &	4081.6584	&	7.60	\\
3798.6337	&	1.82	& \hspace{10mm} &	4084.9154	&	0.48	\\
3799.5965	&	2.78	& \hspace{10mm} &	4086.6439	&	6.64	\\
3803.6372	&	1.51	& \hspace{10mm} &	4090.6314	&	6.14	\\
3805.5997	&	1.28	& \hspace{10mm} &	4093.6188	&	5.04	\\
3808.6153	&	1.60	& \hspace{10mm} &	4094.6146	&	6.86	\\
3810.6021	&	1.82	& \hspace{10mm} &	4095.6139	&	3.13	\\
3997.8792	&	3.14	& \hspace{10mm} &	4100.5988	&	6.88	\\
4000.9181	&	2.24	& \hspace{10mm} &	4101.5946	&	5.05	\\
4003.8632	&	3.61	& \hspace{10mm} &	4102.5953	&	6.81	\\
4005.8595	&	3.77	& \hspace{10mm} &	4103.5930	&	6.81	\\
4010.8476	&	3.61	& \hspace{10mm} &	4104.5869	&	6.96	\\
4023.9142	&	2.54	& \hspace{10mm} &	4105.5876	&	6.87	\\
4024.8081	&	5.19	& \hspace{10mm} &	4108.5755	&	6.95	\\
4025.8054	&	5.21	& \hspace{10mm} &	4116.5703	&	6.48	\\
4028.7963	&	5.51	& \hspace{10mm} &	4117.5707	&	6.48	\\
4029.7944	&	5.51	& \hspace{10mm} &	4124.5788	&	5.75	\\
4030.7936	&	5.52	& \hspace{10mm} &	4127.5763	&	5.58	\\
4031.7896	&	5.67	& \hspace{10mm} &	4128.5810	&	5.39	\\
4035.7800	&	6.00	& \hspace{10mm} &	4134.5845	&	4.94	\\
4038.8807	&	3.66	& \hspace{10mm} &	4135.6248	&	3.96	\\
4048.7434	&	2.33	& \hspace{10mm} &	4136.5857	&	4.84	\\
4049.7408	&	2.33	& \hspace{10mm} &	4140.5878	&	4.55	\\
4050.7387	&	2.33	& \hspace{10mm} &	4143.6649	&	2.39	\\
4054.7279	&	2.33	& \hspace{10mm} &	4146.6432	&	2.82	\\
4055.7273	&	2.17	& \hspace{10mm} &	4147.6938	&	1.44	\\
4056.7237	&	2.33	& \hspace{10mm} &	4152.5903	&	3.67	\\
4057.7199	&	2.33	& \hspace{10mm} &	4153.5906	&	2.24	\\
4058.7168	&	2.33	& \hspace{10mm} &	4154.5913	&	3.51	\\
4059.7159	&	2.17	& \hspace{10mm} &	4156.5923	&	3.35	\\
4060.7150	&	2.17	& \hspace{10mm} &	4157.5928	&	3.27	\\
4061.7102	&	2.33	& \hspace{10mm} &	4158.5935	&	3.24	\\
4062.7057	&	2.33	& \hspace{10mm} &	4160.5943	&	3.03	\\
4064.7010	&	2.33	& \hspace{10mm} &	4161.6335	&	2.08	\\
4067.6962	&	2.18	& \hspace{10mm} &	4169.6321	&	1.60	\\
4069.6876	&	2.33	& \hspace{10mm} &	4170.5992	&	2.24	\\
4070.6872	&	2.18	& \hspace{10mm} &	4171.6000	&	2.24	\\
4071.6826	&	2.34	& \hspace{10mm} &	4172.6004	&	2.08	\\
4072.6826	&	2.24	& \hspace{10mm} &	4173.6007	&	2.08	\\
4074.6737	&	2.34	& \hspace{10mm} &	4174.6013	&	1.92	\\
4077.6648	&	2.34	& \hspace{10mm} &	4175.6017	&	1.92	\\
4178.6033	&	1.60	& \hspace{10mm} &			&		\\
\enddata
\end{deluxetable}
\clearpage
\begin{deluxetable}{lrcrr}
\tablecolumns{5}
\tablewidth{0pt}
\tablenum{2}
\tablecaption{Detected pulsation frequencies of EE~Cam\label{TABfreqs}}
\tablehead{
 \colhead{}		&
 \colhead{Frequency}	&
 \colhead{Detection}	&
 \colhead{Amplitude}	&
 \colhead{Notes}	\\
 \colhead{}		&
 \colhead{}		&
 \colhead{Significance}	&
 \colhead{$y$ filter}	&
 \colhead{}		\\
 \colhead{}		&
 \colhead{cd$^{-1}$}	&
 \colhead{Ampl. S/N}	&
 \colhead{mag}		&
 \colhead{}
}
\startdata
	&		&		& $\pm$ 0.0005	& 		\\
f1	&  4.934	& 97 		& 0.0360 	&		\\
f2	&  5.214	& 51		& 0.0195 	&		\\
f3	&  9.840	& 16		& 0.0065 	&		\\
f4	&  8.333	& 15		& 0.0061	&		\\
f5	&  8.457	& 13		& 0.0049 	&		\\
f6	&  4.937	&  9.8 		& 0.0036 	&		\\
f7	& 10.147	&  8.7 		& 0.0036	& =f1+f2	\\
f8	&  7.905	&  7.9 		& 0.0032 	&		\\
f9	&  4.765	&  7.8 		& 0.0029	&		\\
f10	& 10.869	&  6.5 		& 0.0027	&		\\
f11	&  6.205	&  6.2 		& 0.0024	&		\\
f12	&  7.263	&  5.9 		& 0.0024	&		\\
f13	&  9.548	&  5.2 		& 0.0021	&		\\
f14	&  7.665	&  4.9 		& 0.0019 	&		\\
f15	& 10.319	&  4.6 		& 0.0019	&		\\
\enddata
\tablecomments{A detection is considered significant if the\\
amplitude signal/noise ratio $\geq$ 4.00 (Breger et al. 1993).\\
This is similar to a power signal/noise ratio of $\sim$ 12.6.\\
The noise was calculated over 2 cd$^{-1}$ ranges.}
\end{deluxetable}

\end{document}